\documentclass[journal,twoside]{IEEEtran}
\usepackage{amsmath}
\usepackage{amsfonts}
\usepackage{amssymb}
\usepackage{makeidx}
\usepackage{graphicx}
\usepackage{epstopdf}
\usepackage{cite}
\usepackage{color}

\ifCLASSINFOpdf

 \newtheorem{theorem}{Theorem}

 \newtheorem{lemma}{Lemma}
\else

\fi

\newcommand{\yu}{\mathbf{y}_{\text{u}}}
\newcommand{\pu}{P_{\text{u}}}
\newcommand{\pr}{P_{\text{r}}}
\newcommand{\pp}{P_{\text{p}}}

\newcommand{\nr}{\mathbf{n}_{\text{r}}}

\newcommand{\G}{\mathbf{G}}
\newcommand{\D}{\mathbf{D}}
\newcommand{\Gh}{\hat{\mathbf{G}}}
\newcommand{\g}{\mathbf{g}}
\newcommand{\gh}{\hat{\mathbf{g}}}

\newcommand{\yr}{\mathbf{y}_{\text{R}}}

\newcommand{\B}[1]{\mathbf{#1}}

\begin{document}

\title{On the Performance of Zero-Forcing Processing in Multi-Way Massive MIMO Relay Networks}
\vspace{-1cm}
\author{{Chung Duc Ho, \emph{Student Member, IEEE}, Hien Quoc Ngo, \emph{Member, IEEE}, Michail Matthaiou, \emph{Senior Member, IEEE}, and Trung Q. Duong, \emph{Senior Member, IEEE}}
\vspace{-1cm}

\thanks{
        Manuscript received November 26, 2016; accepted January 2 2017. The associate editor coordinating the review of this paper and approving it for publication was C. Masouros.
        The authors are with the School of Electronics, Electrical Engineering and Computer Science, Queen’s University Belfast, Belfast, U.K. (\IEEEauthorblockA{\IEEEauthorrefmark{0}e-mail:\{choduc01, m.matthaiou, trung.q.duong\}@qub.ac.uk, hien.ngo@liu.se}). H. Q. Ngo is also with the Department of Electrical Engineering (ISY), Link\"{o}ping University,  Link\"{o}ping, Sweden. 
            
        This work was supported by project no. 3811/QD-UBND, Binh Duong government, Vietnam. The work of H. Q. Ngo was supported by the Swedish Research Council (VR) and ELLIIT. The work of M. Matthaiou was supported in part by the EPSRC under grant EP/P000673/1. The work of T. Q. Duong was supported by the U.K. Royal Academy of Engineering Research Fellowship under Grant RF1415$\backslash$14$\backslash$22.
}

}
\markboth{IEEE COMMUNICATIONS LETTERS, ACCEPTED FOR PUBLICATION}
        {On the Performance of Zero-Forcing Processing in Multi-Way Massive MIMO Relay Networks}

\maketitle
\renewcommand{\baselinestretch}{0.93} \normalsize
\begin{abstract}
We consider a multi-way massive multiple-input multiple-output  relay network with zero-forcing processing at the relay. By taking into account the time-division duplex protocol with channel estimation, we derive an  analytical approximation  of  the  spectral efficiency. This approximation is very tight and simple which enables  us  to  analyze the system performance, as well as, to compare the spectral efficiency with zero-forcing and maximum-ratio processing. Our results show that by using a very large number of relay antennas and with the zero-forcing technique, we can simultaneously serve many active users in the same time-frequency resource, each with high spectral efficiency.

\end{abstract}

%
%

\IEEEpeerreviewmaketitle
\vspace{-0.4cm}
\section{Introduction}
Multi-way relay networks are relevant for many applications, such as data transfer in multimedia teleconference and data exchange between sensor nodes and data fusion centers in wireless communications \cite{gunduz2013multiway}. Due to the multiplexing gain, the spectral efficiency of multi-way relay networks is much larger than that of two-way or one-way relay networks. Therefore, during the past years, multi-way relay networks have attracted considerable research interest \cite{amah2009non}. On a parallel avenue, massive multiple-input multiple-output (MIMO) has also attracted a significant amount of research interest from both  academia and industry \cite{Eri:13:MCOM}. In massive MIMO, hundreds of antennas are deployed at the base station to serve simultaneously tens of users. With simple linear processing techniques, such as maximum-ratio (MR) or zero-forcing (ZF) processing, massive MIMO can offer huge spectral and energy efficiency \cite{ngo2013energy}. Thus, massive MIMO  combined  with  multi-way relaying technique  is  a  strong  candidate  for the next-generation wireless communication systems.

Recently, there have been some works in multi-way massive MIMO  relay systems  \cite{amarasuriya2013multi,amarasuriya2014multi}.  These systems can offer all benefits of both massive MIMO and multi-way relaying technologies, and hence, are expected to offer very high spectral efficiency. In particular, in \cite{amarasuriya2013multi}, the authors show that by using very large antenna arrays at the relay together with ZF processing, the system performance can improve significantly. Furthermore,  \cite{amarasuriya2014multi}  shows that the transmit power of each user and/or the relay can be made inversely proportional to the number of relay antennas, while maintaining a required quality of service. However, these works assume perfect channel state information (CSI) at the relay and users. In practice, especially in massive MIMO systems, the impact of channel estimation should be taken into consideration. In \cite{HNMD:16:ICC}, the authors analyze the performance of multi-way massive MIMO systems with imperfect CSI and MR processing at the relay. To the best of the authors' knowledge, there is no work on ZF processing with imperfect CSI in literature, partially due to the difficulty in manipulating products of Wishart matrices.

In this paper, we investigate a multi-way massive MIMO relay network with ZF processing and imperfect CSI. The relay estimates the channels via uplink pilots and the minimum mean square error (MMSE) scheme. Then, it uses these channel estimates and the ZF technique to combine and beamform the signals to all users. We derive an approximate closed-form expression for the spectral efficiency. This approximation is very tight and enables us to further analyze the performance of the considered system.

${\textit{Notation}:}$ The superscripts $(\cdot)^T$, $(\cdot)^*$, and $(\cdot)^H$ stand for the transpose, conjugate, and Hermitian, respectively. The notations $\mathbb{E}\{\cdot\}$ and $\mathbb{V}$ar$\{\cdot\}$ are the expectation and the variance operators, respectively. Furthermore, $\B{[A]}_{k}$ or $\B{a}_{k}$ denotes the $k$-th column of matrix $\B{A}$.

\vspace{-0.2cm}
\section{Multi-Way Massive MIMO Relay Model}


We consider a multi-way relaying massive MIMO network which includes one relay station and $K$ users.\footnote{
It would be more practical to consider multi-cell setups.
Unfortunately, if we consider multi-cell setups, the system model becomes too complicated
to analyze. Note that our results can be regarded as an upper bound of
what is actually achieved in multi-cell setups. If a pilot reuse scheme is applied, then this upper bound is very tight \cite{marzetta2016fundamentals}.} The relay station is equipped with $M$ antennas, and each user has a single antenna $(M>K)$. In this system, each user wants to communicate with $K-1$ other users with the aid of the relay. We assume that the direct links (user-to-user links) are absent due to large path loss and/or severe shadowing. 

Let $\G\in \mathbb{C}^{M\times K}$ be the channel matrix from the $K$ users to the relay, which includes the small-scale fading and the large-scale fading and is modeled as
\begin{align}
\G=\B{H}\B{D}^{1/2},
\end{align}
where $\B{H}\sim \mathcal{CN}(0,\B{I}_M)$ represents the small-scale fading,  and $\D\in\mathbb{C}^{K\times K}$ is a diagonal matrix containing the large-scale fading coefficients whose $k$-th diagonal element is denoted by $\beta_{k}$.

The transmission protocol is the same as the one in \cite{HNMD:16:ICC}. More precisely, the data exchange between all the $K$ users is done via time-division duplex (TDD) operation. With TDD operation, each coherence interval is divided into three phases: channel estimation, multiple-access, and  broadcast  phases. 

\IEEEpeerreviewmaketitle
\vspace{-0.3cm}
\subsection{Channel Estimation Phase}
All the $K$ users simultaneously send pilot sequences to the relay. The relay then estimates the channels to all users through receiving pilots. Let $T$ and $\tau$ be the lengths of each coherence interval and the training duration (in symbols), respectively, with $T>\tau$. We assume that the pilots used by the $K$ users are pairwisely orthogonal. This requires $\tau\geq K$. We denote by $\pp$ the normalized transmit signal-to-noise ratio (SNR) per pilot symbol. Then, the MMSE channel estimate of $\B{G}$ can be represented as \cite{HNMD:16:ICC}
\begin{equation}
\hat{\G}= \G - \B{\tt{\B{E}}},
\end{equation}
where $\B{E}$ is the estimation error matrix, which is independent of $\Gh$. Furthermore, $\hat{\G}\sim \mathcal{CN}(0,\hat{\B{D}})$ and $\B{E}\sim \mathcal{CN}(0,\B{D}_{\text{E}})$, where $\hat{\B{D}}$ and $\B{D}_{\text{E}}$ are diagonal matrices whose $(k,k)$-th elements are
$\sigma^2_k=\frac{\tau\pp\beta^2_k}{\tau\pp\beta_k+1},
$
and
$
\sigma^2_{e,k}=\beta_{k}-\sigma^2_{k}
$, respectively.

\vspace{-0.2cm}
\subsection{Multiple-Access Phase}
After sending the pilot sequences' phase, all the users simultaneously send their data to the relay. Let $x_k$, where $\mathbb{E}\left\{\left|x_{k}\right|^2\right\}=1$, is the signal transmitted from the $k$-th user. Then, the relay sees 
\begin{align}\label{eq:yr}
\B{y}_\text{R}=\sqrt{\pu}\B{G}\B{x}+\nr,
\end{align}
where $\pu$ is the normalized transmit SNR, $\B{x}\triangleq~[x_{1},\ldots, x_{K}]^T$, and $\nr\sim\mathcal{CN}(0,\B{I}_M)$ is the AWGN vector at the relay. Then, the relay uses the channel estimate and  ZF technique to combine the received signals from all $M$ antennas as 
\begin{align}\label{eq:tilde_yr}
\tilde{\B{y}}_{\text{R}}=\B{W}^T\yr,
\end{align}
where $\B{W}^T$ is the ZF receiver given by \cite{ngo2013energy}  
\begin{align}
\B{W}^T&=\left(\Gh^H\Gh\right)^{-1}\Gh^H.\label{eq:W^T_ZF}
\end{align}

\IEEEpeerreviewmaketitle
\vspace{-0.3cm}
\subsection{Broadcast Phase}
To send all signals to $K$ users, the relay spends $K-1$ time-slots. In the $t$-th time-slot, the relay aims to send $x_{k+t}$ to user $k$ (if $k+t>K$, then $x_{k+t}$ is set to be $x_{k+t-K}$). 
Thus, the transmit signal vector at the relay for the $t$-th time-slot is
\begin{align}\label{eq:S_R1}
\B{s}_\text{R}^{(t)}
&=
\sqrt{\alpha^{(t)}}\B{A}\B{\Pi}^{(t)}\tilde{\B{y}}_{\text{R}}, \quad t=1, 2,\dots, K,
\end{align}
where $\B{\Pi}^{(t)}\in\mathbb{C}^{K\times K}$ is the permutation matrix at the $t$-th time-slot  given by \cite[Eq.~(17)]{amarasuriya2013multi}, $\B{A}$ is the ZF precoding matrix expressed as 
\vspace{-0.2cm}
\begin{align}
\B{A}=\Gh^*\left(\Gh^T\Gh^*\right)^{-1},\label{eq:A_ZF}
\end{align}
\vspace{-0.2cm}
and  $\alpha^{(t)}$ is chosen to satisfy the power constraint at the relay,
\begin{equation}\label{eq:pc_relay}
\mathbb{E}\left\{\left\|  \B{s}_\text{R}^{(t)} \right\|^2\right\}=\pr.
\end{equation}
Denote
$\B{B}^{(t)}\triangleq\B{A}\B{\Pi}^{(t)}\B{W}^T\G$, and
$\B{C}^{(t)}\triangleq\B{A}\B{\Pi}^{(t)}\B{W}^T$.
Then $\B{s}_\text{R}^{(t)}$ in \eqref{eq:S_R1}  can be rewritten as
\begin{align}\label{eq:S_R}
\B{s}_\text{R}^{(t)}
&=
\sqrt{\alpha^{(t)}\pu}\B{B}^{(t)}\B{x}+\sqrt{\alpha^{(t)}}\B{C}^{(t)}\B{n}_\text{r}.
\end{align}
Plugging \eqref{eq:S_R} into \eqref{eq:pc_relay}, we have
\begin{align}\label{eq:alpha_t_zf}
\alpha^{(t)}=\frac{\pr}{\pu{\tt{Q}}_{\tt{1}}^{(t)}+\pu{\tt{Q}}_{\tt{2}}^{(t)}+{\tt{Q}}_{\tt{3}}^{(t)}},
\end{align}
where
\begin{align}
{\tt{Q}}_{\tt{1}}^{(t)}
&\triangleq\mathbb{E}\left\{\text{Tr}\left[\left(\B{A}\B{\Pi}^{(t)}\B{W}^T\Gh\right)\left(\B{A}\B{\Pi}^{(t)}\B{W}^T\Gh\right)^H\right]\right\},\label{eq:Q1}\\
{\tt{Q}}_{\tt{2}}^{(t)}
&\triangleq\mathbb{E}\left\{\text{Tr}\left[\left(\B{A}\B{\Pi}^{(t)}\B{W}^T\B{E}\right)\left(\B{A}\B{\Pi}^{(t)}\B{W}^T\B{E}\right)^H\right]\right\},\label{eq:Q2}\\
{\tt{Q}}_{\tt{3}}^{(t)}
&\triangleq\mathbb{E}\left\{\text{Tr}\left[\left(\B{A}\B{\Pi}^{(t)}\B{W}^T\right)\left(\B{A}\B{\Pi}^{(t)}\B{W}^T\right)^H\right]\right\}.\label{eq:Q3}
\end{align}
With the transmitted signal given in \eqref{eq:S_R}, the $K$ users  receive
\begin{align}\label{eq:y_u_1}
{\yu}^{(t)}
&=
\G^T\B{s}^{(t)}_\text{R}+\B{n}^{(t)}_\text{u}\nonumber\\
&=
\sqrt{\alpha^{(t)}\pu}\G^T\B{B}^{(t)}\B{x}+\sqrt{\alpha^{(t)}}\G^T\B{C}^{(t)}\B{n}_\text{r}+\B{n}^{(t)}_\text{u},
\end{align}
where  $\B{n}_{\text{u}}\sim\mathcal{CN}(0,\B{I}_K)$ is the AWGN vector at the users. 

\section{Spectral Efficiency Analysis}

We derive a  closed-form expression for the spectral efficiency of the transmission in the first time-slot.  The same analysis can be done for other time-slots. Note that, hereafter, we set $k+1=1$ if $k=K$ and set $k-1=K$ if $k=1$.

By using the bounding technique in \cite{ngo2014multipair}, the received signal  at the $k$-th user  $y_{\text{u},k}^{(1)}$ is expressed as:
\begin{align}\label{eq:y_u_t}
y_{\text{u},k}^{(1)}
&=
\underbrace{\sqrt{\alpha^{(1)}\pu}\mathbb{E}\left\{\g^T_{k}\B{b}^{(1)}_{k+1}\right\}x_{k+1}}_{\text{desired signal}}+\underbrace{\tilde{N}^{(1)}_{k}}_{\text{effective noise}},
\end{align}
where 
\begin{align}\label{eq:effective_noise}
&\tilde{N}^{(1)}_{k}
\triangleq
\sqrt{\alpha^{(1)}\pu}\left(\g^T_{k}\B{b}^{(1)}_{k+1}-\mathbb{E}\left\{\g^T_{k}\B{b}^{(1)}_{k+1}\right\}\right)x_{k+1}\nonumber\\
&+
\sqrt{\alpha^{(1)}\pu}\!\!\!\!\sum_{\overset{i=1}{i\neq (k+1)}}^{K}\!\!\!\!\g^{T}_{k}\B{b}^{(1)}_{i}x_{i}
+
\sqrt{\alpha^{(1)}}\g^T_{k}\B{C}^{(1)}\B{n}_\text{r}+n^{(1)}_{\text{u},k}.
\end{align}
The worst-case Gaussian noise yields an achievable spectral efficiency 
for the $k$-th user, which is given as
\begin{align}\label{eq:SE_k}
{\tt{SE}}^{(1)}_{k}
	\!=\!
	\left(\!\frac{T\!-\!\tau}{T}\!\right)\!\!\left(\!\frac{K\!-\!1}{K}\!\right)\!\log_2\!\left(\!1+\frac{\alpha^{(1)}\pu\left|\mathbb{E}\left\{\g^T_{k}\B{b}^{(1)}_{k+1}\right\}\right|^2}{\mathbb{V}\text{ar}\left(\tilde{N}^{(1)}_{k}\right)}\!\right).
\end{align}
To derive the spectral efficiency in closed-form, we need to compute $\mathbb{E}\left\{\g^T_{k}\B{b}^{(1)}_{k+1}\right\}$ and $\mathbb{V}\text{ar}\left(\tilde{N}^{(1)}_{k}\right)$. From  the independence between  $\Gh$ and $\B{E}$, we have
\begin{align}
\mathbb{E}\left\{\g^T_{k}\B{b}^{(1)}_{k+1}\right\}
 &=
 \mathbb{E}\left\{\gh^T_{k}\B{A}\B{\Pi}^{(1)}\B{W}^T\gh_{k+1}\right\} =1.
\end{align}
Since $\tilde{N}^{(1)}_{k}$ has a complicated form which includes matrix inversions and multiplications of Wishart matrices, we cannot obtain an exact closed-form of  $\mathbb{V}\text{ar}\left(\tilde{N}^{(1)}_{k}\right)$. However, thanks to the law of large numbers (for large $M$), we can obtain the following approximation.
\begin{theorem}\label{Th:theorem_1}
	As $M\to \infty$, the spectral efficiency \eqref{eq:SE_k} can be approximated as
	\setcounter{equation}{31}
	\begin{align}\label{eq:theo11}
	&{\tt{SE}}^{(1)}_{k}
	\!\rightarrow \!\left(\!\frac{T\!-\!\tau}{T}\!\right)\!\!\left(\!\frac{K\!-\!1}{K}\!\right)\!\log_2\!\!\left(\!\!1\!+\!\frac{\alpha^{(1)}\pu}{\alpha^{(1)}\!\pu\!\!\sum\limits_{i=1}^{K} \!\mathcal{I}_{k,i}\!+\!\alpha^{(1)}\! \mathcal{J}_{k}\!+\!1}\!\!\right),
	\end{align}
	where
	\begin{align}
	&\alpha^{(1)}\triangleq\frac{M(M-K)\pr}{M\pu\sum_{k=1}^{K}\frac{1}{\sigma^{2}_{k}}+\pu\sum_{k=1}^{K}\sigma^2_{e,k}\varrho+\varrho},\label{eq:alpha_zf}\\
	&\mathcal{I}_{k,i}\!\triangleq\!\frac{M\sigma^2_{i-1}\sigma^2_{e,i}\!+\!M\sigma^2_{k+1}\sigma^2_{e,k}\!+\!\sigma^2_{k+1}\sigma^2_{i-1}\sigma^2_{e,k}\sigma^2_{e,i}\varrho}{M(M-K)\sigma^2_{i-1}\sigma^2_{k+1}},\label{eq:I22}\\
	&\mathcal{J}_{k}\triangleq\frac{M+\sigma^2_{k+1}\sigma^2_{e,k}\varrho}{M(M-K)\sigma^2_{k+1}},\ \text{and} \ \ \label{eq:AN_kk}
	\varrho\triangleq \sum_{k'=1}^{K}\frac{1}{\sigma^{2}_{k'}\sigma^2_{k'+1}}.
	\end{align}

\begin{IEEEproof}
	See Appendix~\ref{AP:Theorem_1}.
\end{IEEEproof}
	
\end{theorem}


\IEEEpeerreviewmaketitle
\vspace{-0.3cm}
\section{Numerical Results}

We consider the sum spectral efficiency as our performance metric. The sum spectral efficiency is defined as
\IEEEpeerreviewmaketitle
\vspace{-0.1cm}
\begin{align}
{\tt{SE}_{sum}}=\sum_{k=1}^{K}{\tt{SE}}^{(1)}_{k} \ \ \ \text{bit/s/Hz}.
\end{align}

\begin{figure}[t]
	\centerline{\includegraphics[scale=0.24]{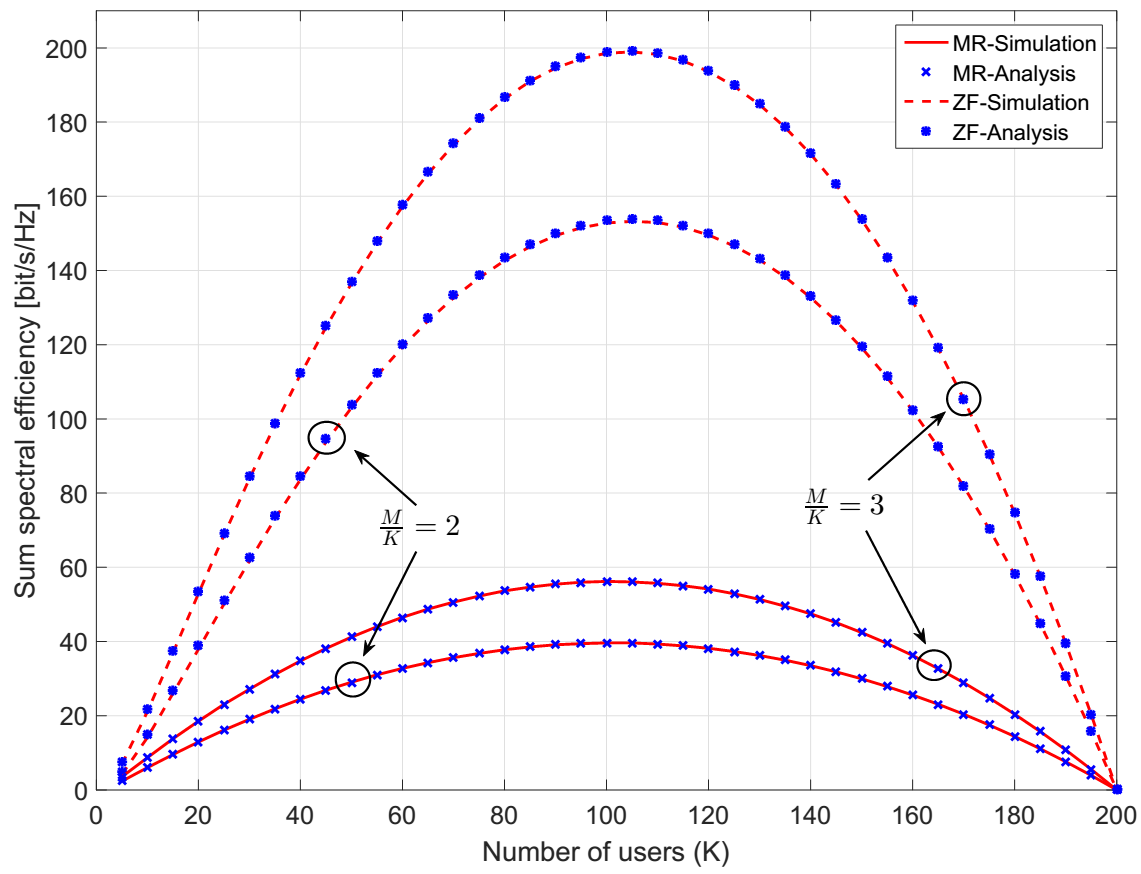}}
	\caption{
		Sum spectral efficiency versus the number of users $K$. We choose $T=200$, $\pu=\pp=0$ dB, $\pr=10$ dB, and $\beta_{k}=1$.
	}\label{fig:ratio_M_over_K}
\end{figure}

For the first example, we assume that $\beta_{k}=1$, and choose $T=200, \pu=\pp=0$ dB, $\pr=10$ dB. Figure~\ref{fig:ratio_M_over_K} compares the performance of multi-way massive MIMO systems for ZF and MR processing with different $K$, while the ratio $M/K$ is kept fixed. For MR processing, we used the results in \cite[Eq.~(26)]{HNMD:16:ICC}.  Clearly, the simulated spectral efficiency and the approximate one  match perfectly. At a small $K$ (low inter-user interference) and large $K$ (large channel estimation overhead),  the spectral efficiencies of ZF and MR processing are comparable. However, when $K=20$--$180$, ZF significantly outperforms MR processing. Interestingly, regardless of the ratio $M/K$, the sum spectral efficiency is maximum when $K$ is around $100$. Furthermore, when $M/K$ increases, the inter-user interference reduces, and hence, the sum spectral efficiency increases.

\begin{figure}[t]
	\centerline{\includegraphics[scale=0.25]{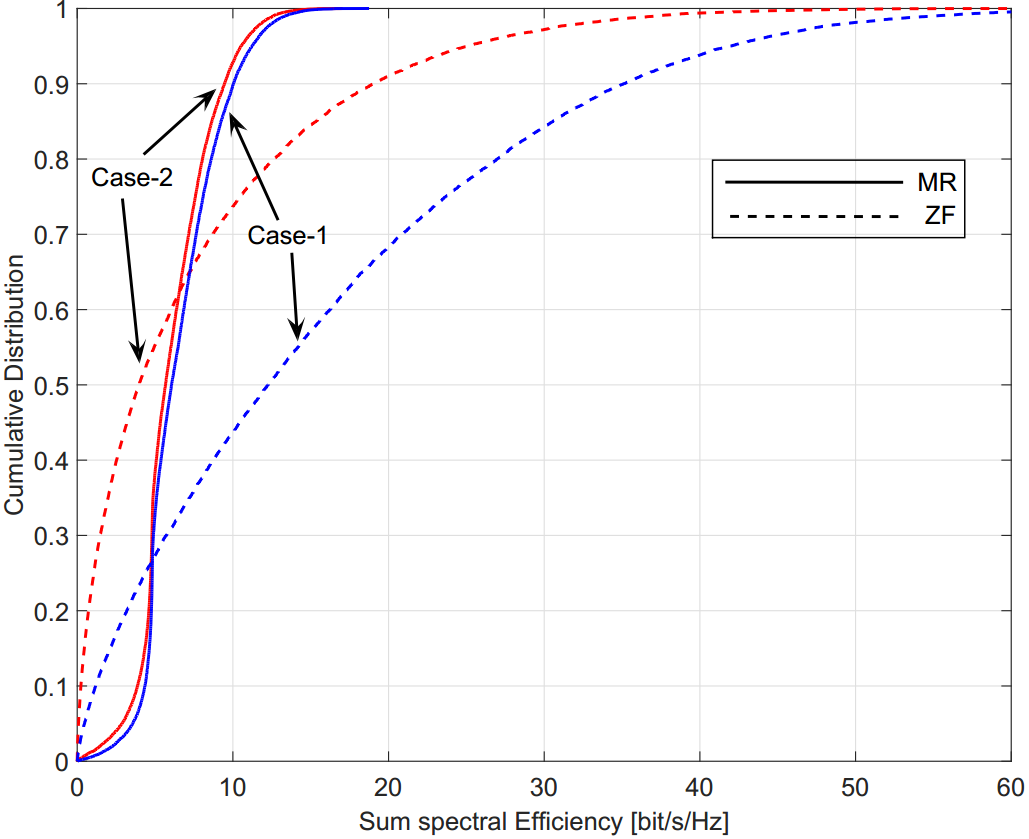}}
	\caption{
		Cumulative distribution of the sum spectral efficiency. Here $M=100$, $K=20$. 
	}\label{fig:cdf}
\end{figure}

We next consider a more practical scenario where users  are  located uniformly at random inside a disk with the diameter of $1000$~m. The large-scale fading is modelled as
$
\beta_{k}=\frac{z_{k}}{1+(d_{k})^\nu},
$
where $z_{k}$ is the log-normal random variable with standard deviation of $8$~dB, $\nu=4$ denotes the path-loss exponent, and $d_k$ is the distance between the $k$-th user and the relay. Furthermore, the normalized transmit SNRs $P_\text{r}$, $P_\text{u}$ and $P_\text{p}$ can be calculated by dividing these powers by the noise power $N_0$. In this example, we choose $N_0 = -120$~dB. We consider 2 cases: Case-1 corresponds to ($P_\text{u}=P_\text{p} =0.2$~W,  $P_\text{r}=1$~W), and Case-2 corresponds to ($P_\text{u}=P_\text{p} =0.1$~W,  $P_\text{r}=0.5$~W). Figure~\ref{fig:cdf} shows the cumulative distribution of the sum spectral efficiency   for  ZF and MR processing. We can see that, at high transmit power (Case-1),  the spectral efficiency of ZF processing is higher than  the one of MR processing and vice versa at low transmit power. Furthermore, compared with  MR processing, the spectral efficiency of ZF processing is less concentrated around its median.

\vspace{-0.2cm}

\IEEEpeerreviewmaketitle
\vspace{-0.1cm}
\section{Conclusion}
We have investigated a multi-way massive MIMO relay network with ZF processing and imperfect CSI. We derived a new tractable approximate closed-form expression for the spectral efficiency. For a large number of relay antennas, the inter-user interference and noise reduces significantly, and hence, the system can deliver a substantial sum spectral efficiency. Furthermore, we showed that, for most of the cases (particularly at high SNRs), ZF processing offers a higher spectral efficiency than MR processing  does.

\vspace{-0.2cm}
\section{Appendices}

\subsection{Preliminary Results}
\begin{lemma}\label{LM:lemma1}
		Let $\B{X}\in\mathbb{C}^{M\times K}$, $M>K$. Each row of $\B{X}$ is $\mathcal{CN}(0,\B{D})$, where $\B{D}$ is a diagonal matrix. Furthermore, let $\hat{\B{D}}\in \mathbb{C}^{K\times K}$ be another diagonal matrix. Then, we have 
		\begin{align}\label{eq:lemma1}
		\mathbb{E}\left\{\text{Tr}\left[\hat{\B{D}}\left(\B{X}^H\B{X}\right)^{-1}\right]\right\}
		&=\frac{1}{M-K}\sum_{k=1}^{K}\frac{[\hat{\B{D}}]_{kk}}{[{\B{D}}]_{kk}}.		
		\end{align}			
\begin{IEEEproof}
By expressing $\text{Tr}\left[\hat{\B{D}}\left(\B{X}^H\B{X}\right)^{-1}\right] = \sum_{k=1}^{K}\frac{[\hat{\B{D}}]_{kk}}{[{\B{D}}]_{kk}} [\mathcal{W}^{-1}]_{kk}$, where $\mathcal{W}$ is a $K\times K$ central Wishart matrix of $M$ degrees of freedom, and using \cite[Lemma~2.10]{tulino2004random}, we obtain \eqref{eq:lemma1}.
\end{IEEEproof}
\end{lemma}

\begin{lemma}\label{LM:lemma2}
		Let $\B{A}\in\mathbb{C}^{M\times M}$, and $\B{x}\sim\mathcal{CN}(0,\B{I}_{M})$. Then,
		\begin{align}\label{eq:lemma2}
		\mathbb{E}\left\{\left|\B{x}^T\B{A}\B{x}\right|^2\right\}=\text{Tr}\left(\B{A}\B{A}^H\right)+\text{Tr}\left(\B{A}\B{A}^*\right).
		\end{align}	
\begin{IEEEproof}To obtain \eqref{eq:lemma2}, we first express $\B{x}^T\B{A}\B{x} $ as $\sum_{m=1}^M\sum_{m'=1}^M a_{mm'}x_m x_{m'}$, and then use the identities $\mathbb{E}\left\{|x_m|^4\right\}=2$ and $\mathbb{E}\left\{|x_m x_{m'}|^2\right\} =1$, for $m\neq m'$.
\end{IEEEproof}

\end{lemma}

\subsection{Proof of Theorem \ref{Th:theorem_1}}\label{AP:Theorem_1}

\subsubsection{Derivation of $\alpha^{(1)}$}
From \eqref{eq:alpha_t_zf}, to compute $\alpha^{(1)}$ we need to compute ${\tt{Q}}_{\tt{1}}^{(1)}$, ${\tt{Q}}_{\tt{2}}^{(1)}$, and ${\tt{Q}}_{\tt{3}}^{(1)}$.
The substitution of \eqref{eq:W^T_ZF} and \eqref{eq:A_ZF} into \eqref{eq:Q1}  yields
\begin{align}\label{eq:appQ1}
{\tt{Q}}_{\tt{1}}^{(1)}
&=\mathbb{E}\left\{\text{Tr}\left(\Gh^T\Gh^*\right)^{-1}\right\}=\frac{1}{M-K}\sum_{k=1}^{K}\frac{1}{\sigma^{2}_{k}},
\end{align}
where in the last equality we have used Lemma~\ref{LM:lemma1}.

To compute ${\tt{Q}}_{\tt{2}}^{(1)}$, we substitute \eqref{eq:W^T_ZF} and \eqref{eq:A_ZF} into \eqref{eq:Q2}  to obtain
\begin{align*}
{\tt{Q}}_{\tt{2}}^{(1)}
&\!=\!\sum_{k=1}^{K}\!\sigma^2_{e,k}\mathbb{E}\left\{\!\text{Tr}\!\!\left[\B{\Pi}^{(1)}\!\!\left(\Gh^H\Gh\right)^{-1}\!\!\left(\B{\Pi}^{(1)}\right)^H\!\!\left(\Gh^T\Gh^*\right)^{-1}\!\right]\!\right\}.
\end{align*}
From the law of large numbers, we have that $\Gh^H\Gh\rightarrow M\hat{\B{D}}$, and hence, ${\tt{Q}}_{\tt{2}}$ can be approximated as
\begin{align}\label{eq:appQ2}
{\tt{Q}}_{\tt{2}}^{(1)}
&\!\rightarrow\!\sum_{k=1}^{K}\!\sigma^2_{e,k}\mathbb{E}\!\left\{\text{Tr}\!\left[\!\B{\Pi}^{(1)}\!\left(M\hat{\B{D}}\right)^{-1}\!\!\left(\B{\Pi}^{(1)}\!\right)^H\!\!\left(\Gh^T\Gh^*\right)^{-1}\!\right]\!\right\}\nonumber\\
&=\frac{\varrho}{M(M-K)}\sum_{k=1}^{K}\sigma^2_{e,k},
\end{align}
where again we have used Lemma~\ref{LM:lemma1} to obtain the last equality.

Similarly, we obtain
\begin{align}\label{eq:appQ3}
{\tt{Q}}_{\tt{3}}^{(1)}
	\rightarrow
	\frac{\varrho}{M(M-K)}.
\end{align}

Substituting \eqref{eq:appQ1}, \eqref{eq:appQ2}, and \eqref{eq:appQ3} into \eqref{eq:alpha_t_zf}, we obtain \eqref{eq:alpha_zf}.

\subsubsection{Derivation of $\mathbb{V}\mathrm{ar}\left(\tilde{N}^{(1)}_{k}\right) $}
From \eqref{eq:effective_noise}, we have
\begin{align}\label{eq:app_var}
	&\mathbb{V}\text{ar}\left(\tilde{N}^{(1)}_{k}\right) 
	= 
	\alpha^{(1)}\pu\mathbb{V}\text{ar}\left(\g^T_{k}\B{b}^{(1)}_{k+1}\right) + \alpha^{(1)}\pu\mathbb{E}\left\{\left|\g^{T}_{k}\B{b}^{(1)}_{k}\right|^2\right\}\nonumber\\
	&+\alpha^{(1)}\pu\!\!\!\!\!\!\!\sum_{\overset{i=1}{i\neq (k, k+1)}}^{K}\!\!\!\!\!\!\!\mathbb{E}\!\left\{\!\left|\g^{T}_{k}\B{b}^{(1)}_{i}\right|^2\!\right\}
	\!+\!\alpha^{(1)}\mathbb{E}\!\left\{\!\left\|\g^T_{k}\B{C}^{(1)}\right\|^2\!\right\}+\!1.
\end{align}

\begin{itemize}
\item[a)]  Compute $\mathbb{V}\text{ar}\left(\g^T_{k}\B{b}^{(1)}_{k+1}\right)$: By expressing the true channel as the sum of the channel estimate plus the channel estimation error, we obtain
\begin{align}
\mathbb{V}\text{ar}\left(\g^T_{k}\B{b}^{(1)}_{k+1}\right)
&=  {\tt{V}}_1 + {\tt{V}}_2 + {\tt{V}}_3,
 \label{eq:var}
\end{align}
where 
\vspace{-0.2cm}
\begin{align}
{\tt{V}}_1\triangleq\mathbb{E}\left\{\left|\gh^T_{k}\B{A}\B{\Pi}^{(1)}\B{W}^T\B{e}_{k+1}\right|^2\right\},\\
{\tt{V}}_2\triangleq\mathbb{E}\left\{\left|\B{e}^T_{k}\B{A}\B{\Pi}^{(1)}\B{W}^T\gh_{k+1}\right|^2\right\},\\
{\tt{V}}_3\triangleq\mathbb{E}\left\{\left|\B{e}^T_{k}\B{A}\B{\Pi}^{(1)}\B{W}^T\B{e}_{k+1}\right|^2\right\}.
\end{align}
The term ${\tt{V}}_1$ can be computed as
\begin{align*}
&{\tt{V}}_1
=\sigma^2_{e,k+1}\mathbb{E}\left\{\left\|\gh^T_{k}\B{A}\B{\Pi}^{(1)}\B{W}^T\right\|^2\right\}\nonumber\\
&=\sigma^2_{e,k+1}\mathbb{E}\left\{\!\left[\left(\Gh^H\Gh\right)^{-1}\right]_{k+1,k+1}\!\right\}\!=\!\frac{\sigma^2_{e,k+1}}{(M\!-\!K)\sigma^2_{k+1}}.
\end{align*}
Similarly, we obtain
$
{\tt{V}}_2
=\frac{\sigma^2_{e,k}}{(M-K)\sigma^2_{k}},
$
and
$
{\tt{V}}_3
=\frac{\sigma^2_{e,k}\sigma^2_{e,k+1}}{M(M-K)}\varrho
$. Therefore, 
	\begin{align}\label{eq:appvar_zf}
	&\mathbb{V}\text{ar}\left(\g^T_{k}\B{b}^{(1)}_{k+1}\right)
	= \mathcal{I}_{k,k+1}.
	\end{align}

\item[b)] Compute $\mathbb{E}\left\{\left|\g^{T}_{k}\B{b}^{(1)}_{k}\right|^2\right\}$: By expressing $\B{g}_k=\hat{\B{g}}_k + \B{e}_k$, and using the fact that 
$\gh^T_{k}\B{A}\B{\Pi}^{(1)}\B{W}^T\gh_{k} = 0,$
we get
\begin{align}
\mathbb{E}\left\{\left|\g^{T}_{k}\B{b}^{(1)}_{k}\right|^2\right\}
&=\mathbb{E}\left\{\left|{\tt{I}}_{1}+{\tt{I}}_{2}+{\tt{I}}_{3}\right|^2\right\},\label{eq:I_1}
\end{align}
where
$
{\tt{I}}_{1}\triangleq\gh^T_{k}\B{A}\B{\Pi}^{(1)}\B{W}^T\B{e}_{k}$,
${\tt{I}}_{2}\triangleq\B{e}^T_{k}\B{A}\B{\Pi}^{(1)}\B{W}^T\gh_{k}$, and
${\tt{I}}_{3}\triangleq\B{e}^T_{k}\B{A}\B{\Pi}^{(1)}\B{W}^T\B{e}_{k}$. Since ${\tt{I}}_{1}$, ${\tt{I}}_{2}$, and ${\tt{I}}_{3}$ are mutually uncorrelated, we obtain
\begin{align}
\mathbb{E}\!\left\{\!\left|\g^{T}_{k}\B{b}^{(1)}_{k}\right|^2\!\right\}
&\!=\!\mathbb{E}\left\{\!\left|{\tt{I}}_{1}\right|^2\!\right\}\!\!+\mathbb{E}\!\left\{\!\left|{\tt{I}}_{2}\right|^2\!\right\}\!+\!\mathbb{E}\!\left\{\!\left|{\tt{I}}_{3}\right|^2\!\right\}.\label{eq:I1}
\end{align}

Similarly to the derivation of ${\tt V}_1$, we have
\begin{align}
\mathbb{E}\!\left\{\!\left|{\tt{I}}_{1}\right|^2\!\right\}\!=\!
\frac{\beta_{k}\!-\!\sigma^2_{k}}{(M\!-\!K)\sigma^2_{k+1}}, 
\mathbb{E}\!\left\{\!\left|{\tt{I}}_{2}\right|^2\!\right\}=
\frac{\beta_{k}\!-\!\sigma^2_{k}}{(M\!-\!K)\sigma^2_{k-1}}.\label{eq:I_13}
\end{align}

Next, we compute $\mathbb{E}\left\{\left|{\tt{I}}_{3}\right|^2\right\}$. Using \eqref{eq:lemma2} from Lemma~\ref{LM:lemma2}, and the law of large numbers, we obtain
\begin{align}
&\mathbb{E}\left\{\left|{\tt{I}}_{3}\right|^2\right\}\nonumber\\
&\!=\!\sigma^4_{e,k}\mathbb{E}\left\{\text{Tr}\left[\B{\Pi}^{(1)}\left(\Gh^H\Gh\right)^{-1}\!\!\left(\B{\Pi}^{(1)}\right)^H\!\!\left(\Gh^T\Gh^*\right)^{-1}\right]\right\}\nonumber\\
&\!+\!\sigma^4_{e,k}\mathbb{E}\left\{\text{Tr}\left[\B{\Pi}^{(1)}\left(\Gh^H\Gh\right)^{-1}\!\!\left(\B{\Pi}^{(1)}\right)^*\!\!\left(\Gh^T\Gh^*\right)^{-1}\right]\right\}\nonumber\\
&\rightarrow\frac{\sigma^4_{e,k}\varrho}{M(M-K)}.\label{eq:I14_re}
\end{align}

Substituting \eqref{eq:I_13}, and \eqref{eq:I14_re} into \eqref{eq:I1}, we get 
	\begin{align}\label{eq:I1_rr}
	\mathbb{E}\left\{\left|\g^{T}_{k}\B{b}^{(1)}_{k}\right|^2\right\}
	\rightarrow
	\mathcal{I}_{k,k}.
	\end{align}

\item[c)] Compute $\mathbb{E}\left\{\left|\g^{T}_{k}\B{b}^{(1)}_{i}\right|^2\right\}$, where $i\neq k, k+1$: Following a similar methodology as in the derivation of $\mathbb{E}\left\{\left|\g^{T}_{k}\B{b}^{(1)}_{k}\right|^2\right\}$, we obtain
	\begin{align}\label{eq:I22}
	&\mathbb{E}\left\{\left|\g^{T}_{k}\B{b}^{(1)}_{i}\right|^2\right\}
	\rightarrow\mathcal{I}_{k,i}.
	\end{align}

\item[d)] Compute $\mathbb{E}\left\{\left\|\g^T_{k}\B{C}^{(1)}\right\|^2\right\}$: By replacing $\g_{k}$ with $\hat{\g}_{k} +\B{e}_k$ together with Lemma~\ref{LM:lemma1} and the law of large numbers, as in the derivation of $\mathbb{E}\left\{\left|\g^{T}_{k}\B{b}^{(1)}_{k}\right|^2\right\}$, we obtain
\begin{align}
		\mathbb{E}\left\{\left\|\g^T_{k}\B{C}^{(1)}\right\|^2\right\}=\frac{M+\sigma^2_{k+1}\sigma^2_{e,k}\varrho}{M(M-K)\sigma^2_{k+1}}\label{eq:AN_kk}.
		\end{align}

\end{itemize}

Substituting \eqref{eq:appvar_zf}, \eqref{eq:I1_rr}, \eqref{eq:I22}, and \eqref{eq:AN_kk} into \eqref{eq:app_var} yields \eqref{eq:theo11}.

%

\bibliographystyle{IEEEtran}
\bibliography{bibfile}

\end{document}